\documentclass[twocolumn,showpacs,english,prl,superscriptaddress,preprintnumbers,amsmath,amssymb,floatfix]{revtex4-1}

\usepackage[T1]{fontenc}
\usepackage[latin9]{inputenc}
\usepackage{epsfig,psfrag}
\usepackage{dcolumn}
\usepackage{bm}
\usepackage{graphicx}
\usepackage{color}
\usepackage{esint}
\usepackage{babel}
\usepackage{amsfonts}
\usepackage{enumerate}

\begin{document}

\title{ Multi-Terminal Coulomb-Majorana Junction }

\author{Alexander Altland}
\affiliation{ Institut f\"ur Theoretische Physik, 
Universit\"at zu K\"oln, Z\"ulpicher Str.~77, D-50937 K\"oln, Germany }

\author{Reinhold Egger}
\affiliation{ Institut f\"ur Theoretische Physik, 
Heinrich-Heine-Universit\"at, D-40225  D\"usseldorf, Germany }

\date{\today}

\begin{abstract}
We study multiple helical nanowires in proximity to a common 
mesoscopic superconducting island, where Majorana fermion bound states
are formed.  We show that a weak finite charging energy of the center
island may dramatically affect the low-energy behavior of the
system. While for strong charging interactions, the junction decouples
the connecting wires, interactions \textit{lower} than a non-universal
threshold may trigger the flow towards an exotic Kondo fixed point. In
either case, the ideally Andreev reflecting fixed point characteristic
for infinite capacitance (grounded) devices gets destabilized by
interactions. 
\end{abstract}

\maketitle

\textit{Introduction.---}Electronic transport through 
topological insulators \cite{hasan} or superconductors \cite{qizhang} 
has come into the limelight of condensed-matter physics over the
past few years.  In particular, understanding 
the physics of localized Majorana bound states,
generically expected near the ends of one-dimensional (1D) 
topological superconductor wires \cite{kitaev,fukane,sato}, 
is crucial for exploiting the non-Abelian statistics of Majorana fermions 
in topological quantum computation schemes \cite{alicea,beenakker,karsten}.  
When a grounded Majorana nanowire --- such as InSb and InAs wires 
subject to a Zeeman field and proximity coupling to an $s$-wave
superconductor~\cite{lutchyn,oreg} --- is weakly contacted by a normal metal,
the presence of a Majorana state should reflect in a 
conductance peak~\cite{demler,nilsson,law,flensberg,wimmer}, 
and signatures of this type were indeed observed 
experimentally~\cite{leo,exp1,exp2,exp3}.

\begin{figure}
\centering
\includegraphics[width=8cm]{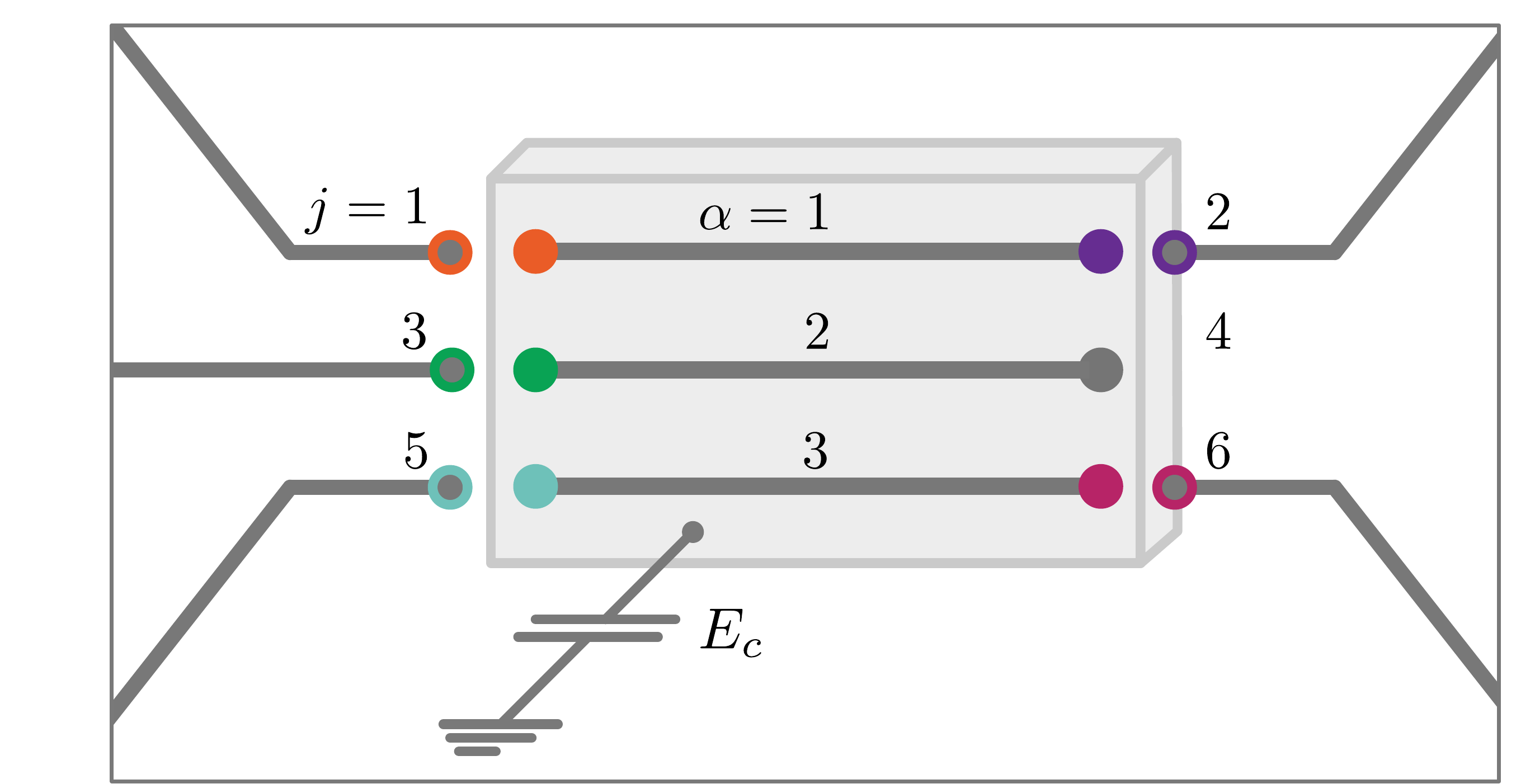}
\caption{\label{fig1} (Color online) 
Schematic setup of the multi-terminal Coulomb-Majorana junction.
$N$ helical nanowires (here, $N=3$)
connect to a floating mesoscopic superconductor (shown as gray box)
with charging energy $E_c$.  Majorana fermions $\gamma_j$ 
(filled circles) exist near the ends of proximity-coupled 
wire segments.  The remaining wire parts (away from the island)
act as LL leads connected to the island by
tunnel couplings $t_j$.  Majorana fermions representing
Klein factors are shown as open circles.  The case of $M<2N$ leads
(here, $M=5$) is included by putting one or several $t_j=0$.
Direct tunneling between the $\gamma_j$ is neglected.  }
\end{figure}

In this paper, we discuss novel transport phenomena caused by Coulomb
interactions in devices comprising Majorana wires contacted to leads.
Our setup is sketched in Fig.~\ref{fig1}: For $N$ nanowires
proximity-coupled to the same mesoscopic superconducting island ('dot'), there
are $2N$ Majorana fermions, $\gamma_j=\gamma_j^\dagger$, with
anticommutator $\{\gamma_j,\gamma_{k}\} =\delta_{jk}$.  The island
connects to the $j$th lead by tunneling through the Majorana state
$\gamma_j$ with coupling strength $t_j$; all other coupling mechanisms
are irrelevant \cite{fidkowski}.  Coulomb interactions affect the
system in two distinct ways: (i) The combination of repulsive
interactions, spin-orbit coupling and Zeeman field is expected to turn
each of the $M\leq 2N$ \textit{connecting leads} into a spinless
(helical) Luttinger liquid (LL) characterized by an interaction
constant $g<1$~\cite{gogolin,delft,alex2}. (ii) Weak intra-island
interactions do not compromise the integrity of individual
Majoranas~\cite{loss1,stoud,eran}. However, they introduce
correlations \textit{between} these states, and thereby correlations
between the connecting wires. It stands to reason that the option to
generate inter-wire coupling on mesoscopic scales, i.e., independently
of microscopic single-particle tunneling between distinct
terminals~\cite{alicea1,halperin}, offers new perspectives for quantum
computation (and other) applications.

One of our main observations is that the charging energy of the
island, $E_c$, no matter how weak, always destabilizes the fully
Andreev reflecting fixed point characterizing the grounded system.
The ensuing consequences can be conveniently described in a 
picture wherein tunneling events from and into lead $j$ 
correspond to particles and antiparticles carrying a flavor index $j$.  
At time scales $\tau\lesssim E_c^{-1}$, these 'particles' are 
asymptotically free and the corresponding tunnel amplitudes, $t_j$, 
scale up. At scales $\tau \sim E_c^{-1}$,
this upward renormalization towards a fully Andreev reflecting fixed
point is stopped by the formation of a strong 'confinement force'
between particles and antiparticles, which enforces electro-neutrality
of the dot. At larger scales, the effective degrees of freedom are
charge dipoles, describing near instantaneous (at time scales $\sim
E_c^{-1}$) transmission of charge from lead $j$ to lead $k \not=j$.
The effective dipole coupling strengths, $\lambda{_{jk}}$, are subject
to a competition of downward renormalization due to fluctuations of
individual dipoles, and upward renormalization due to dipole-dipole
interactions. The balance of these two mechanisms defines an
isotropic repulsive fixed point, $\lambda^\ast$, separating a flow
towards the decoupled dot ($\lambda\to 0$) from a flow towards an
exotic Kondo regime ($\lambda\to \infty$), generalizing the $M=3$
topological Kondo effect discussed in Ref.~\cite{beri1}.
Surprisingly, our analysis below shows that for sufficiently
\textit{weak}\ charging energy, the system always flows towards the
strong-coupling regime.  We finally note that in supporting a novel
type of Kondo flow, the present system differs strongly from the
$M=2$ Majorana single-charge transistor~\cite{fu,zazu,msct} as well
as from other types of previously studied LL 
junctions~\cite{nayak,oshikawa,alex}.

\textit{Model.---}We start by deriving the low-energy effective theory
for the setup in Fig.~\ref{fig1}.  The Hamiltonian, $H=H_c+H_t+H_l$,
contains a piece $H_c$ describing the island, the tunnel Hamiltonian
$H_t$, and $H_l$ for the LL leads (we put $\hbar=k_B=1$ below).  For
each nanowire ($\alpha=1,\ldots,N$), we have two spatially well
separated Majorana fermions, $\gamma_{2\alpha-1}$ and
$\gamma_{2\alpha}.$ It is useful to define the nonlocal auxiliary
fermion operators
$d_\alpha=(\gamma_{2\alpha-1}+i\gamma_{2\alpha})/\sqrt{2}$, where the
total Majorana occupation number operator is $\hat n=\sum_\alpha
d^\dagger_\alpha d^{}_\alpha$. Assuming that the proximity gap exceeds
all other energy scales, the island is fully described in terms of the
$\gamma_j$ and the Cooper pair number operator, $\hat N_c$, or
conjugate phase, $\varphi$, where $[\varphi,2\hat N_c]=i$. Since both,
the Majoranas, and the superconducting phase are zero-energy modes,
the island Hamiltonian is solely due to charging effects,
\begin{equation}\label{hc}
H_c = E_c (2\hat N_c+\hat n)^2.
\end{equation}
Here, we have dismissed the optional dependence of $H_c$ on some
background gate voltage as physically insignificant to our 'nearly
open' dot.  The semi-infinite LL leads, with tunnel contacts
at $x=0$, are described by dual bosonic 
fields, $\phi_j(x)$ and $\theta_j(x)$ \cite{gogolin}, where
\begin{equation}\label{hl}
H_l =\frac{v_F}{2\pi} \sum_{j} \int_0^\infty dx \left[
g(\partial_x\phi_j)^2+ g^{-1}(\partial_x\theta_j)^2\right]
\end{equation}
with Fermi velocity $v_F$. The fermion annihilation operator
for a right- or left-mover reads 
$\psi_{j,R/L}(x)= a^{-1/2} \eta_j e^{i(\phi_j\pm \theta_j)}$, 
where $a$ is a short-distance cutoff
and the $\eta_j$ are additional Majorana fermions with $\{\eta_j,
\eta_{k}\}=\delta_{jk}$ \cite{delft,oshikawa}.  
These 'Klein factors' ensure fermion anticommutation relations 
between different leads and play an important role below.
Since $H_l$ describes decoupled leads with perfect normal
reflection at $x=0$, we impose open boundary conditions, 
$\psi_{j,L}(0)=\psi_{j,R}(0)$, pinning all fields $\theta_j(0)$.
A lead fermion operator near the respective tunnel contact is thus 
given by $\Psi_j=a^{-1/2}\eta_j e^{i\phi_j(0)}$.
Finally, the tunnel Hamiltonian reads~\cite{zazu}
\begin{equation}\label{ht}
H_t = \sum_j \tilde t_j
\Psi_j^\dagger \left( d^{}_{\alpha_j} + (-)^{j-1} e^{-2i\varphi} 
d_{\alpha_j}^\dagger \right) + {\rm H.c.},
\end{equation}
where $\tilde t_j= (a/2)^{1/2}t_j$ and $\alpha_j=[j/2]+1$ (cf. Fig. \ref{fig1}.)
The creation of a lead electron lowers the occupation of
the dot either by annihilation  of a $d$-fermion (first term), or by
creation of a $d$-fermion along with annhilation of a Cooper pair
(the second term.) 
 
\textit{Effective phase action.---} We next derive an effective
action, $S$, describing the system in terms of phase-like degree of
freedoms.  Integrating over the harmonic LL bosons away from $x=0$, we
obtain a contribution $S_l= \frac{Tg}{2\pi} \sum_{j,n} |\omega_n|
|\Phi_j(\omega_n)|^2$, describing how fluctuations of
$\Phi_j(\tau)\equiv \phi_j(0,\tau)$ dissipate into lead
excitations ($\omega_n=2\pi n T$ are Matsubara frequencies and $T$
is temperature.)  Next, by virtue of a Hubbard-Stratonovich
transformation to the condensate phase field $\varphi(\tau)$, the
charging term in Eq.~\eqref{hc} leads to the contribution
\begin{equation}\label{sc2}
S_c=  \int_0^{1/T} d\tau \left( \frac{\dot\varphi^2}{4E_c} 
-i \dot\varphi \hat n\right).
\end{equation}
Noting that there also is a free fermion piece, $S_f= \sum_\alpha \int
d\tau \bar d_\alpha \dot d_\alpha + \frac12 \sum_j \int d\tau
\eta_j\dot \eta_j$, we eliminate the $\dot\varphi \hat n$ term in
Eq.~\eqref{sc2} by a gauge transformation, $d_\alpha(\tau) \to
e^{-i\varphi(\tau)} d_\alpha(\tau)$.  Switching back to the language
of Majoranas, $\gamma_{2\alpha-1}= (d_\alpha^{} +
d_\alpha^\dagger)/\sqrt{2}$ and
$\gamma_{2\alpha}=-i(d_\alpha^{}-d^\dagger_{\alpha})/\sqrt{2}$, the
tunnel action now assumes the form
\begin{equation}\label{tunnel}
S_t=\sum_j t_j \int d\tau (-2i\eta_j\gamma_j) 
\sin(\Phi_j+\varphi).
\end{equation}
Crucially, the Klein factor $\eta_j$ and the Majorana $\gamma_j$ can
be fused to form an auxiliary fermion
$f_j=(\eta_j-i\gamma_j)/\sqrt{2}$, where $-2i\eta_j\gamma_j
=2f_j^\dagger f_j^{}-1 \equiv \sigma_j=\pm$.  The resulting fermion
Hamiltonian conserves $f^\dagger_j f_j^{}$, which means that the
functional integral over the $f_j$-fermion reduces to a sum over the
static quantum number $\sigma_j=\pm$ \cite{foot}. The sole effect of
this summation is to eliminate contributions of odd order in the
tunneling amplitudes $t_j$ to  the
functional integral, while even-order contributions remain
unaffected. Keeping only even orders, we will drop the then immaterial
presence of the $\sigma_j$'s throughout.  This Klein-Majorana fusion
procedure implies an enormous technical simplification, since we are
now left with an effective phase action only.  Before writing down
this action, we shift $\Phi_j\to \Phi_j-\varphi$, thus removing
$\varphi$ from the tunnel action \eqref{tunnel}.  Performing the
remaining Gaussian functional integral over $\varphi$, after unitary
transformation to the discrete Fourier basis ($q=0,\ldots,M-1$),
\begin{equation}\label{fourier}
\tilde\Phi_q (\tau) = \frac{1}{\sqrt{M}} 
\sum_{j=1}^{M} e^{i 2\pi  q j/M} \Phi_j (\tau) ,
\end{equation}
we arrive at the effective phase action
\begin{eqnarray}\label{aes1}
S &=& \frac{Tg}{2\pi} \sum_{q,n} \frac{|\omega_n|}{1+\delta_{q,0} 
\ \epsilon/|\omega_n|}  
 \left|\tilde\Phi_q(\omega_n)\right|^2
\\ \nonumber &+&\sum_j t_j \int d\tau \sin\Phi_j(\tau),
\end{eqnarray}
with the energy scale $\epsilon \equiv 2g M E_c/\pi$.
Notice that the charging energy only affects the zero mode 
$\tilde \Phi_0$, which effectively becomes \textit{free}\ at low frequencies, 
$|\omega_n|\ll \epsilon$.

\textit{Particle analogy and renormalization.---}We next discuss the expansion in the tunnel couplings $t_j$. 
A very instructive analogy follows by interpreting 
${\cal O}_j^{\pm} (\tau) = e^{\pm i\Phi_j(\tau)}$
as particles (`quarks') and antiparticles living on
the imaginary-time axis. Our particles carry a
`flavor' index ($j=1,\ldots,M$) and an 
effective charge $-it_j/2$.
We study the properties of the ensuing interacting particle gas by
standard renormalization group (RG) methods \cite{alex2,kogut}.  In an
RG step, we integrate over fast $\Phi_j$ modes with frequencies in the
shell $\Lambda/b<|\omega_n|<\Lambda$, with rescaling parameter $b>1$
and a high-energy cutoff $\Lambda$ of the order of the proximity gap.
The RG step is completed by rescaling all frequencies, $\omega\to
b\omega$, such that $\Lambda$ stays invariant.  For small $t_j$, the
particle density is low and different charges (i.e., the $t_j$)
renormalize independently.  We first RG-integrate out modes in the
shell $\epsilon<|\omega_n|<\Lambda$, where the zero mode
$\tilde\Phi_0$ is not significantly affected by $E_c$, see
Eq.~\eqref{aes1}.  Some algebra yields the net scaling dimension
$1-1/(2g)$ for $t_j$, which are therefore RG-relevant couplings for
$g>1/2$. This signals a flow towards the resonant Andreev reflection
fixed point \cite{fidkowski}.  In the lead-non-interacting case,
$g=1$, the scaling dimension $1/2$ represents the naive dimension of a
fermion-Majorana scattering operator. After the RG integration
over the shell $\epsilon<|\omega|<\Lambda$, the renormalized couplings
are then given by $t_j^{(1)} = t_j (\Lambda/\epsilon)^{1-1/(2g)}$.

\begin{figure}
\centering
\includegraphics[width=8cm]{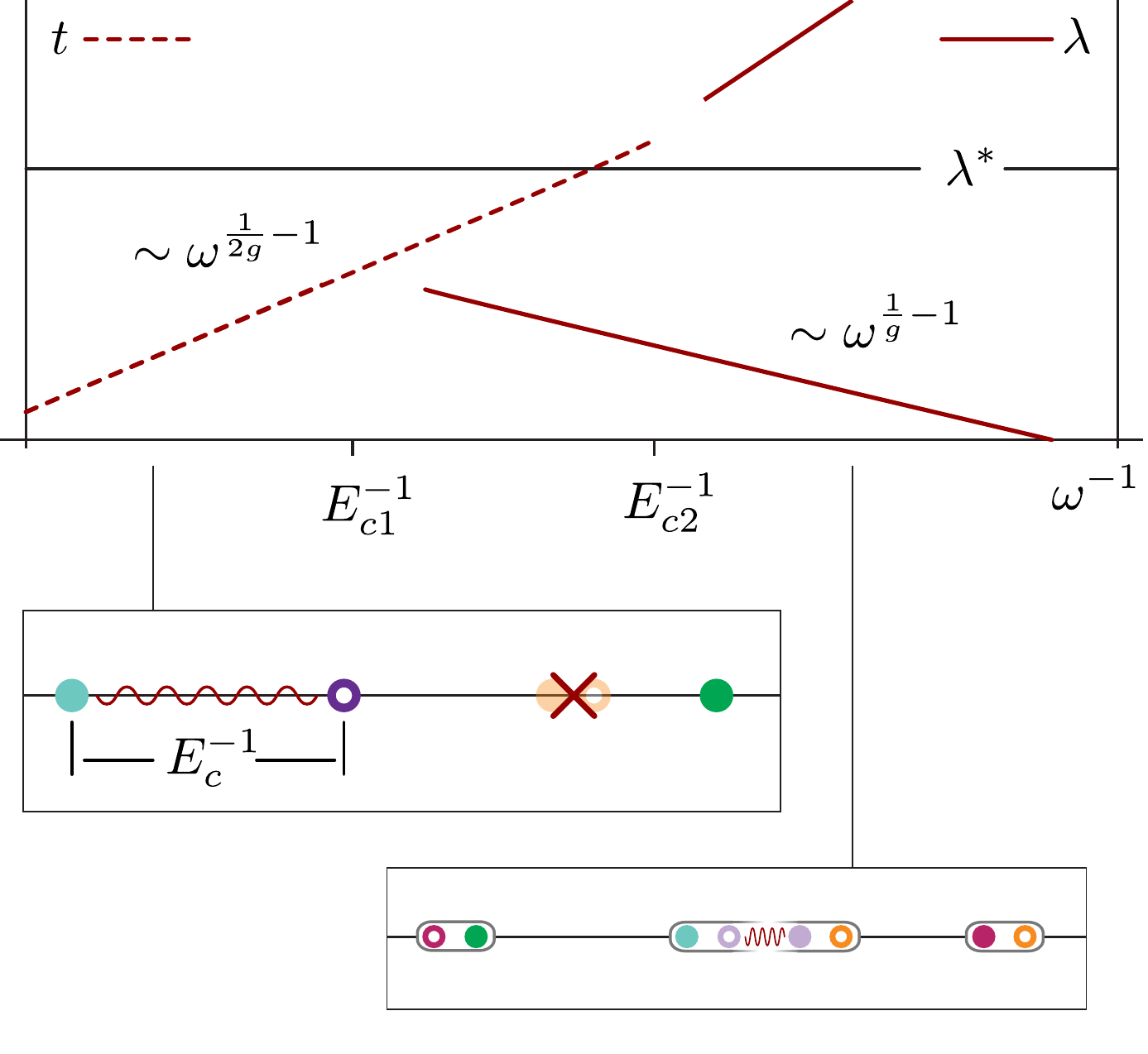}
\caption{\label{fig2} (Color online) Schematic RG flow of typical
couplings, $t$ and $\lambda$, vs the time-like RG parameter
$\omega^{-1}$, and illustration of the system excitations.  At short
times, individual in- and out-tunneling events into different leads
dominate (indicated by dots and circles of different color).  These
`quarks' are asymptotically free, and their fluctuations generate
a power-law increase of $t$.  For $\omega^{-1}\agt E^{-1}_c$, confinement 
leads to bound quark-antiquark dipoles (mesons) for $j\ne k$ (indicated
in the lower inset), or
annihilation for $j=k$ (indicated by 'X' in the  upper inset).
The dipole coupling, $\lambda$, experiences a
downward renormalization due to individual fluctuations and an
upward renormalization due to dipole-dipole fusion events 
(cf.~lower inset).  For sufficiently strong charging, $E_c=E_{c1}$, 
$\lambda$ does not reach the unstable fixed point $\lambda^*$
in Eq.~(\ref{fp}). Then the first mechanism dominates, and 
hence $\lambda$ flows to the decoupled fixed point,  $\lambda=0$,
as $\omega^{-1}\to \infty$.  
For small $E_c=E_{c2}$, however,  the increase of $t$ during the
asymptotic freedom phase brings $\lambda$ beyond the balance point
$\lambda^*$, and the RG flow approaches the strong-coupling
fixed point.}
\end{figure}

Proceeding to lower frequency scales, $|\omega|<\epsilon$, the
charging energy renders the zero-mode action non-dissipative,
$S[\tilde\Phi_0]\simeq \frac{Tg}{2\pi\epsilon} \sum_n \omega_n^2
|\tilde\Phi_0(\omega_n)|^2$, see Eq.~\eqref{aes1}.  The integration
over $\tilde \Phi_0$ generates a strong `confinement' potential linear in the
time separation between 
particle creation, ${\cal
  O}_j^+(\tau) \sim e^{+i\tilde\Phi_0 (\tau) /\sqrt{M}}$, and particle
annihilation events, ${\cal
  O}^-_k(\tau') \sim e^{-i\tilde\Phi_0(\tau')/\sqrt{M}}$ of arbitrary
flavor index $k\not=j$. (For $k=j$,
particle-anti-particle annihilation occurs -- the inconsequential
virtual tunneling of a particle to and fro the same lead,
cf.~Fig.~\ref{fig2} upper inset.) At time
resolutions lower than $E_c^{-1}$, the relevant
excitations of the theory then are quark-antiquark pairs
(`mesons') \cite{foot1},
\begin{eqnarray}\label{had1}
\langle {\cal O}_j^+(\tau) {\cal O}_k^-(\tau')\rangle^{}_0
&\simeq & e^{-\frac{2E_c}{\pi} |\tau-\tau'|}  \ {\cal O}_{jk} (\tau) ,\\
\nonumber
{\cal O}_{jk} (\tau) &= &e^{i\Phi_j(\tau)} e^{-i\Phi_k(\tau)}.
\end{eqnarray}

\textit{RG equations in dipole regime.---} We now consider the
effective low-energy theory at resolutions $E_c\tau>1$, where we are
dealing with a gas of dipoles, ${\cal O}_{jk}$, with couplings
$\lambda_{jk}$, $j\not=k$.  The 'bare value' of these couplings is
given by $\lambda_{jk}^{(1)}\approx t_j^{(1)} t_k^{(1)}/E_c>0$, where
the factor $E_c^{-1}$ is due to a time integration over the distance
between particle and antiparticle. Physically, these couplings bundle
the effect of in-tunneling from lead $j$ into a virtual on-dot state of
longevity $\sim E_c^{-1}$, followed by out-tunneling into lead $k$. 
The effective action describing the system in the dipole regime reads 
\begin{equation} \label{act}
S = \frac{Tg}{2\pi} \sum_n |\omega_n| |\boldsymbol\Phi(\omega_n)|^2
-\sum_{j\ne k} \lambda_{jk}
\int d\tau\ e^{i({\bf e}_j-{\bf e}_k)\cdot \boldsymbol\Phi(\tau)},
\end{equation}
where $\boldsymbol\Phi\equiv(\Phi_1,\dots,\Phi_M)$, and $\mathbf{e}_k$ is a
unit vector in $k$-direction. RG analysis of this action obtains
the flow equations
\begin{equation}\label{rg}
\frac{d\lambda_{jk}}{d\ln b}= \left( 1- g^{-1} \right) \lambda_{jk}
+ \frac{\kappa}{E_c} \sum_{m\neq (j,k)} \lambda_{jm}\lambda_{mk},
\end{equation}
where $\kappa$ is a non-universal constant of order unity.  The first term
describes the standard power-law suppression of the tunneling density
of states in/out of the LL leads \cite{gogolin} and implies a
suppression of the $\lambda_{jk}$.  This term is fought by the
positive second contribution, which describes the effect of dipole
fusion $[(j,m)+(m,k) \to (j,k)]$ by particle/anti-particle
annihilation (cf.~Fig.~\ref{fig2} lower inset.) Equation (\ref{rg})
suggests the existence of an RG-unstable fixed point with isotropic
couplings,
\begin{equation}\label{fp}
\lambda_{jk}=\lambda^* = \frac{g^{-1}-1}{\kappa(M-2)}E_c.
\end{equation}
\noindent When decreasing the effective frequency scale $\omega$, the
couplings, $t$ and $\lambda$, resp., show the RG flow schematically
illustrated in Fig.~\ref{fig2}.  During the initial RG flow (down to
$\omega\approx E_c$), particles are asymptotically free, implying a
power-law increase of a typical tunnel coupling, $t\propto
\omega^{\frac{1}{2g}-1}$, with the time-like variable $\omega^{-1}$.
We now compare the value $\lambda^{(1)}= (t^{(1)})^2/E_c \sim
E_c^{-3+1/g}$, reached at the end of the asymptotic freedom phase
($\omega\approx E_c$), to the fixed-point value $\lambda^*\sim E_c$ in
Eq.~\eqref{fp}.  For sufficiently large $E_c$ \cite{foot2}, the fixed
point cannot be reached, and therefore the $\omega\to 0$ stable fixed
point describes a completely decoupled LL junction.  For sufficiently
small $E_c$, however, the fluctuations generated during the asymptotic
freedom phase tip the balance, $\lambda^{(1)}>\lambda^*$. A
straightforward expansion of Eq.~(\ref{rg}) near $\lambda=\lambda^*$
then shows that the isotropic baseline $\bar \lambda>\lambda^\ast$ of
the couplings flows to large values with dimension $g^{-1}-1>0$, while
deviations between the couplings are RG irrelevant, i.e., the flow is
towards an isotropic configuration $\lambda_{jk}=\bar \lambda$. At a
characteristic 'Kondo temperature', $T_K\sim E_c
\exp\left(-\frac{E_c}{\lambda^{(1)} \kappa (M-2)} \right)$, the
coupling $\bar \lambda$ begins to diverge, and our perturbative
expansion is no longer applicable.

\textit{Strong-coupling fixed point.---} At low frequencies $\omega
\ll T_K$ we are met with a dual picture, where the field vector
$\boldsymbol\Phi$ is pinned to stay close to the minima of the
potential identified by the second term of Eq.~\eqref{act} at
$\lambda_{jk}\simeq \bar \lambda\gg 1$. (Notice, however, that the
zero-mode component $\tilde \Phi_0$ defined through Eq.~\eqref{fourier}
remains free.)  Literally the same problem was studied in
Refs.~\cite{kane,yi}, where it was shown that fluctuations near the
limit $\bar \lambda\to \infty$ -- corresponding to occasional
tunneling events between nearest-neighbor minima -- carry scaling
dimension $2g(M-1)/M$ \cite{yi}. The fixed point
$\bar\lambda\to\infty$ ensuing for $g>M/[2(M-1)]$ was identified with a
multi-channel Kondo fixed point.  However, unlike with previous
proposals, where anisotropy is a relevant perturbation and easily destabilizes
the Kondo fixed point \cite{gogolin}, the present system is robust in that
it flows towards an isotropic configuration. 

\textit{Conductance matrix}.---The two-stage scenario outlined above,
with its branching into either a strong-coupling or a decoupled
low-frequency limit, will bear consequences for all physical
observables.  We here briefly discuss the resulting temperature
dependence of the linear conductance $G_{jk}$ between different leads
$j$ and $k$.  For high $T>E_c$, asymptotic freedom causes the
power-law scaling $G_{jk}\sim T^{-2+1/g}$, while for $T\to 0$, we
either have a vanishing conductance, $G_{jk}\sim T^{-2+2/g}$, if the
decoupled fixed point is approached, or the unitary limit of an
isotropically hybridized junction, $G_{jk}= \frac{2e^2}{h}
\frac{1}{M}$.  (Transport in this case is carried by the fully
isotropic zero mode $\tilde \Phi_0$, which is protected by gauge
invariance.) The above result for the conductance tensor generalizes
the teleportation scenario discussed for $M=2$ by Fu \cite{fu} to
arbitrary channel numbers.  It is worth stressing that for both stable
fixed points (strong coupling or decoupled), resonant Andreev reflection is
unstable, and hence arbitrary $E_c$ destroy the corresponding fixed
point.

\textit{Conclusions.---}In this paper, we have formulated and studied
the problem of junctions of Majorana wires meeting on a
superconducting island with charging energy $E_c$.  We also included
correlations in the leads, since these typically are 1D nanowires
themselves. Our RG analysis reveals that the physics of the system is
determined by asymptotic freedom at short time scales
($\tau<E_c^{-1}$) and confinement at long time scales
($\tau>E_c^{-1}$) corresponding, respectively, to independent dot-lead
tunneling, and virtual co-tunneling. For sufficiently weak $E_c$, we
find that a strong-coupling Kondo fixed point is approached, while
otherwise the junction describes decoupled leads at $T=0$.  These
predictions can be tested using the temperature dependence of the
conductance.

We thank P. Sodano and  A. Zazunov for discussions.
This work was supported by the SFB TR 12 and SPP 1666 of the DFG.
\textit{Note added:} During the preparation of this manuscript, we
learned of independent related work by B{\'e}ri \cite{beri2}.


\begin{thebibliography}{99}
\bibitem{foot} The $\sigma_j=\pm$ correspond to the 
eigenvalues of the Pauli matrix $\sigma_x$ in Ref.~\cite{fidkowski},
which there was found through an elaborate Jordan-Wigner transformation.
\bibitem{foot1}
In the effective action governing the meson (${\cal O}_{jk}$) dynamics
on time scales with $E_c\tau \agt 1$, we can then replace the zero-mode
part by the same dissipative action as for $\tilde\Phi_{q>0}$ without
altering the physics.  While the crossover scale in $S[\tilde\Phi_0]$ is
$\epsilon=2 g M E_c/\pi$, confinement sets in only for
 $|\omega|<E_c$.  For large number of leads $M$, 
this allows for an intermediate frequency
regime which, however, does not qualitatively change our conclusions.
\bibitem{foot2} The threshold value for $E_c$ is non-universal 
and strongly depends on the bare tunnel couplings $t_j$.
\end{thebibliography}
\end{document}